%% file: apssamp.tex
\documentclass[%
preprint,
 amsmath,amssymb,
 aps, physrev,
]{revtex4-2}

\usepackage[title]{appendix}
\usepackage{graphicx}
\usepackage{dcolumn}
\usepackage{bm}

\begin{document}


\title{\textbf{SOT Enabled 3D Magnetic Field Sensor with Low Offset and High Sensitivity} 
}%

\author{Sebastian~Zeilinger}
\email{sebastian.zeilinger@univie.ac.at}
\affiliation{Physics of Functional Materials, Faculty of Physics, University of Vienna, Vienna, Austria}%
\affiliation{Research Platform MMM Mathematics-Magnetism-Materials, University of Vienna, Vienna, Austria}%
\affiliation{Vienna Doctoral School in Physics, University of Vienna, Vienna, Austria}%
\author{Johannes~G\"uttinger}
\affiliation{Infineon Technologies AG, Villach, Austria}
\author{Klemens~Pr\"ugl}%
\affiliation{Infineon Technologies AG, Regensburg, Germany}
\author{Michael~Kirsch}%
\affiliation{Infineon Technologies AG, Regensburg, Germany}
\author{Joshua~M.~Salazar-Mej\'ia}
\affiliation{Vienna Doctoral School in Physics, University of Vienna, Vienna, Austria}%
\author{Sabri~Koraltan}
\affiliation{Institute of Applied Physics, Technische Universität Wien, A-1040, Vienna, Austria}%
\author{Philip~Heinrich}
\affiliation{Physics of Functional Materials, Faculty of Physics, University of Vienna, Vienna, Austria}%
\author{Sophie~Zeilinger}
\affiliation{Faculty Center for Nano Structure Research, Faculty of Physics, University of Vienna, Vienna, Austria}%
\author{Bernd~Aichner}%
\affiliation{Physics of Functional Materials, Faculty of Physics, University of Vienna, Vienna, Austria}%
\author{Florian~Bruckner}
\affiliation{Physics of Functional Materials, Faculty of Physics, University of Vienna, Vienna, Austria}%
\author{Hubert~Br\"uckl}
\affiliation{Department for Integrated Sensor Systems, University for Continuing Education
Krems, 2700 Wiener Neustadt, Austria}
\author{Armin~Satz}
\affiliation{Infineon Technologies AG, Villach, Austria}
\author{Dieter~Suess}%
\affiliation{Physics of Functional Materials, Faculty of Physics, University of Vienna, Vienna, Austria}%
\affiliation{Research Platform MMM Mathematics-Magnetism-Materials, University of Vienna, Vienna, Austria.}%


\date{\today}

\begin{abstract}
In this work we demonstrate a spin orbit torque (SOT)  magnetic field sensor, designed as a Ta/CoFeB/MgO structure, with high sensitivity and capable of active offset compensation in all 3 spatial directions. This is described and verified in experiment and simulation. The measurements of magnetic fields showed an offset of 36, 50 and $37 \, \mathrm{\mu T}$ for x, y and z-fields. Furthermore the sensitivities of these measurements had values of 590, 580 and $490 \, \mathrm{VA^{-1}T^{-1}}$ in x, y and z-direction. In addition the robustness to bias fields is demonstrated via experiments and single spin simulations by applying bias fields in y-direction. Cross sensitivities were further analyzed via single spin simulations performing a parameter sweep of different bias fields in y- and z-direction up to $\pm 1 \, \mathrm{mT}$. Last but not least the extraction of the SOT parameters $\eta_{DL}$ and $\eta_{FL}$ is shown via an optimization of a single spin curve to the experimental measurements.
\end{abstract}

\maketitle


\section{\label{Introduction}Introduction}

Magnetic field sensors are a fundamental part in our modern-day society. They are used in various systems including but not limited to cars, mobile phones and robotics. In multiple applications such as electric current sensors via magnetic fields, it is required to measure the absolute value of the magnetic field with highest accuracy. Here, the zero-field offset is the main limiting factor in state-of-the-art sensors.

On the one hand, a big part of today's magnetic field sensors are Hall based devices that use the spinning current technique to achieve active offset compensation in the micro-Tesla range \cite{SpinningCurrentTechnique, OffsetCompensation}. On the other hand, magnetoresistive (xMR) devices, such as tunneling magnetoresistance sensors, offer very high signal-to-noise ratios but typically lack offset compensation \cite{GiantTunnelingMagnetoresistance, GiantRoomTemp}.

In this work we utilize a modulation principle of the magnetization to reduce the offset. This is realized by spin orbit torque (SOT) that is highly energy efficient and allows for integration into standard CMOS techniques. The SOT effect is under intense focus due to the potential applications in memory and sensor applications \cite{EmergingSpintronics, Current_Induced_SOT, RoadmapOfSOT}. Recently effort is performed to utilize the current induced SOT effect for magnetic field sensing devices \cite{sabrispaper, MagneticAngularSensor, SOTDeviceSensing3DFields}. We present a sensor based on this effect, designed with active offset compensation. In contrast to Refs. \cite{SOTDeviceSensing3DFields} and \cite{SOTLShaped3DSensor} our sensor is also able to measure z-fields offset free, which enables a true 3D sensor using only one sensitive element. Offset compensation in the z-direction is achieved through the spinning current technique, commonly used in conventional Hall devices \cite{Patent, Munter_spinning_current_tech}.

\section{\label{Device_geometry_and_theory}Theory}
\begin{figure}[ht]
    \centering
    \includegraphics[width=\linewidth]{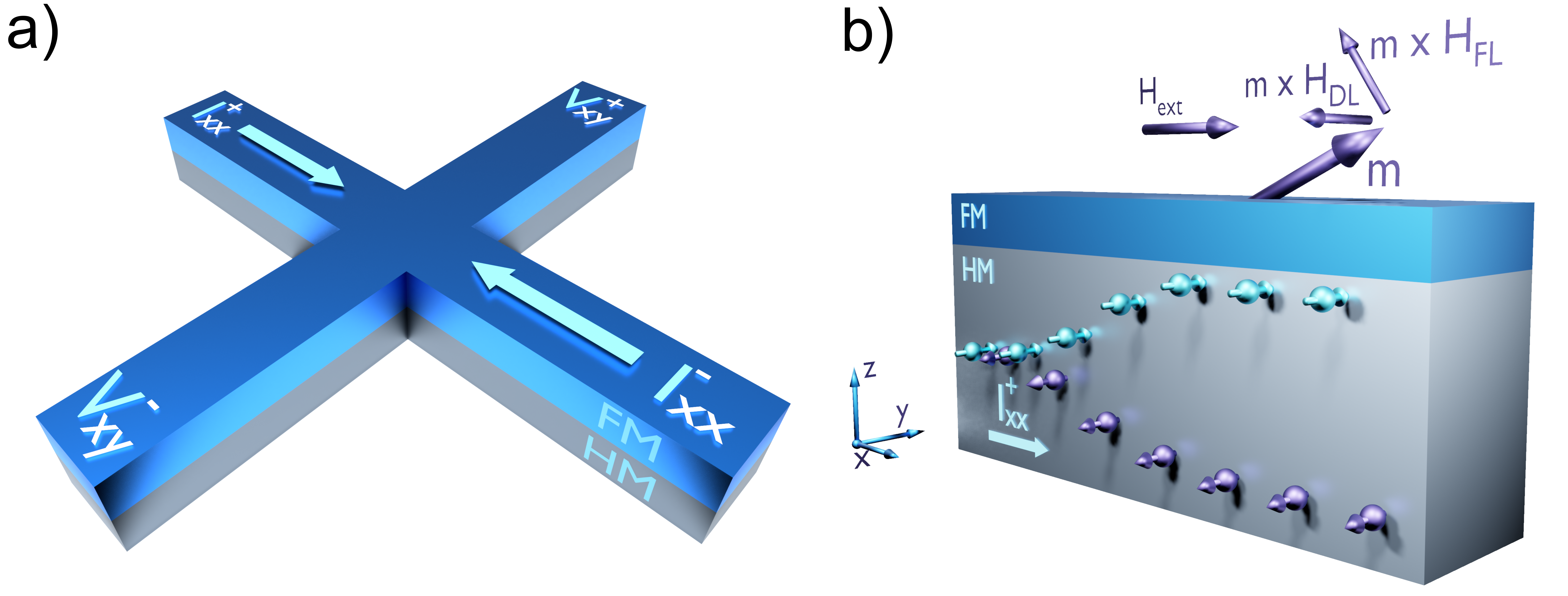}
    \caption{\textbf{a} The analyzed structure is designed as a cross like structure, with a ferromagnetic layer on top of a heavy metal layer. \textbf{b} Spin separation via the Spin Hall effect (SHE). Spin polarized current acts like a torque, through the components $\bm{m} \times \bm{H}_{DL}$ and $\bm{m} \times\bm{H}_{FL}$, on the magnetization of the FM layer under the influence of an external magnetic field ($\bm{H}_{ext}$).}
    \label{cartoon_figure}
\end{figure}

The geometry of the studied device is conceived as a cross like structure with an overall length of $10 \, \mathrm{\mu m}$ and a width of the individual arms of $2 \, \mathrm{\mu m}$ (see Fig. \ref{cartoon_figure}a). The sensor is composed of a ferromagnetic (FM) layer of CoFeB ($1\,\mathrm{nm}$) deposited on a heavy metal (HM) layer of Ta ($6\,\mathrm{nm}$). A MgO ($1.5\,\mathrm{nm}$) / Ta ($5\,\mathrm{nm}$) heterostructure is subsequently deposited over the active layers, although this is not depicted in Fig. \ref{cartoon_figure}a. This design enables the measurement of the perpendicular voltage ($V_{xy}$) in response to an applied current ($I_{xx}$)(see Fig. \ref{cartoon_figure}a).

In order to describe the sensor principle we use the Landau-Lifshitz-Gilbert (LLG) equation augmented with the SOT torque.

\begin{equation}
    \partial _t\bm{m} = -\gamma\bm{m}\times \bm{H}^{\mathrm{eff}}+\alpha\bm{m}\times \partial _t\bm{m}+\bm{T}
    \label{extendedLLG_equation}
\end{equation}

Here $\gamma$ is the gyromagnetic ratio, $\alpha$ the Gilbert damping factor and $\bm{H}^{\mathrm{eff}}$ the effective magnetic field. The torque term $\bm{T}$ \cite{TorqueTerm},

\begin{equation}
    \bm{T}=\gamma H_{FL}\bm{m}\times \bm{p} + \gamma H_{DL}\bm{m}\times \left( \bm{m}\times \bm{p} \right)
    \label{torque_terms_equation}
\end{equation}

is orthogonal to the magnetization and can be expressed as a field-like (FL) part (first term) and a damping-like (DL) part (second term). The spin polarization is expressed with $\bm{p}$, which points in the positive y-direction ($\bm{p}= \left( 0, 1, 0 \right)^\mathrm{T}$) (see Fig. \ref{cartoon_figure}b). $H_{DL}$ and $H_{FL}$ represent the amplitudes of the field components corresponding to the two distinct parts, which can be expressed over the SOT coefficients $\eta_{FL}$ and $\eta_{DL}$ with the following equation \cite{sabrispaper}.

\begin{equation}
    H_{DL/FL} = p_f \cdot \eta _{DL/FL}
    \label{analytical_HDL_Dieter_equation}
\end{equation}

with 

\begin{equation}
    p_f = \frac{{{j_e}\hbar }}{{2e{\mu _0}t{M_s}}}
    \label{pf_factor_Dieter_equation}
\end{equation}

Here $j_e$ is the electrical current density flowing through the HM layer, $t$ the thickness of the FM layer and $M_s$ the saturation magnetization of the FM layer.

The torque term Eq. \eqref{torque_terms_equation} can be inserted in Eq. \eqref{extendedLLG_equation} as field contributions ($\bm{H}_{FL}=H_{FL}\bm{p}$ and $\bm{H}_{DL}=H_{DL}\bm{m}\times\bm{p}$, visible in Fig. \ref{cartoon_figure}b), which act on the magnetization of the FM layer \cite{extendedLLG}.

\begin{equation}
     \partial _t\bm{m} =-\gamma\bm{m}\times \left( \bm{H}^{\mathrm{eff}}- H_{FL} \bm{p} - H_{DL}\bm{m}\times \bm{p}  \right)+\alpha\bm{m}\times \partial _t\bm{m}
     \label{effective_field_equation}
 \end{equation}
 
In systems with low Gilbert damping and with magnetization initially along x due to an external magnetic field, the field-like torque acts like a field along z, pushing the magnetization out of the x–y plane and initiating precession. The damping-like torque pushes the magnetization toward –y, which in our case is opposite to the spin polarization, as illustrated in Fig. \ref{cartoon_figure}b. If the current direction is flipped in the other direction, $H_{DL}$ and $H_{FL}$ change signs leading to opposite out-of-plane (OOP) magnetization. This can be described by analytically solving Eq.\eqref{effective_field_equation} in the linear regime, assuming small fields in the $x$- and/or $z$-directions ($H_x$, $H_z$), and vanishing field in the $y$-direction ($H_y = 0$). The detailed calculation can be seen in Appendix \ref{section_appendix_A}, which leads to 

\begin{equation}
    m_{z,I\pm} =  \frac{\mp H_x \left| p_f \right| \eta_{DL}  + H_z (\left| p_f \right| \eta_{FL} )}{{\left( \left| p_f \right| \eta_{DL} \right)}^2 + (\left| p_f \right| \eta_{FL} ) (\left| p_f \right| \eta_{FL} - H_k )}
    \label{analyitcal_equation_mz_final}
\end{equation}

Here $H_k$ is the amplitude of the anisotropy field pointing in the z-direction. It can be seen if the z-magnetization ($m_{z,I-}$) with negative current is subtracted from the z-magnetization with positive current ($m_{z,I+}$) the response is only sensitive to $H_x$ fields in first order, canceling any cross sensitivities. The signal corresponding to external x-fields ($S_x$) therefore takes the form

\begin{equation}
   S_{x} \left( H_x \right) = \frac{R_s}{2} \cdot \left( m_{z,I+} - m_{z,I-} \right) = -R_s\frac{\eta_{DL} H_x}{-\eta_{FL} H_k + (\eta_{DL}^2 + \eta_{FL}^2) \left| p_f \right|}
   \label{analytical_equation_Sx}
\end{equation}

Here $R_s$ is the anomalous Hall coefficient, used to convert the equation into units of $\Omega$ (see Sec.~\ref{Determination_of_R_s}). This factor was introduced to compare simulation data with measured data. Alternatively, if the magnetizations from the two cases are added, the $H_x$ component is canceled, leaving the system sensitive primarily to external $H_z$ fields in first order. This yields the signal for z-fields ($S_z$), which allows the measurement of external fields without sensitivities in other directions.

\begin{equation}
    S_{z} \left( H_z \right) = \frac{R_s}{2} \cdot \left(m_{z,I+} + m_{z,I-} \right)= R_s\frac{\eta_{FL} H_z}{-\eta_{FL} H_k + (\eta_{DL}^2 + \eta_{FL}^2) \left| p_f \right|}
    \label{analytical_equation_Sz}
\end{equation}

\section{Measurement of 3D magnetic fields}
\label{measurement_of_3D_fields}

\begin{figure}[ht]
    \centering
    \includegraphics[width=\linewidth]{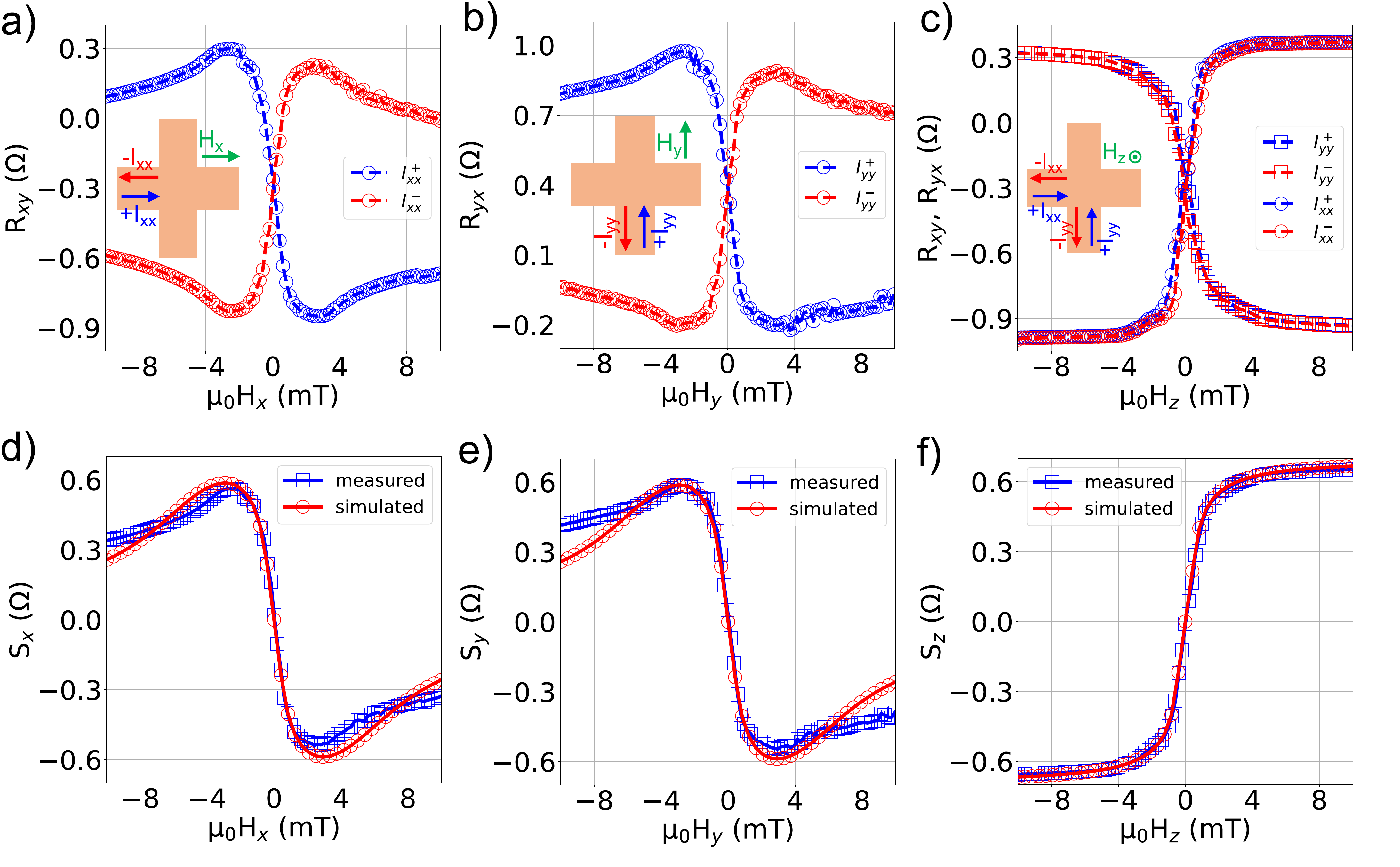}
    \caption{\textbf{a} Measured $R_{xy}$ for $2.7 \, \mathrm{mA}$ current strength under external $H_x$ fields for $I_{xx}^+$ (blue dashed curve with circles) and $I_{xx}^-$ (red dashed curve with circles). \textbf{b} Measured $R_{yx}$ for $2.7 \, \mathrm{mA}$ current strength under external $H_y$ fields. \textbf{c} Measured $R_{xy}$ and $R_{yx}$ for $2.7 \, \mathrm{mA}$ current strength under external $H_z$ fields for $I_{xx}^\pm$ (dashed curves with circles) and $I_{yy}^\pm$ (dashed curves with rectangles). \textbf{d} The sensor response (blue curve with squares) for $2.7 \, \mathrm{mA}$ current strength and single spin simulation results (red curve with circles) under external $H_x$ fields. \textbf{e} The same plot but for an external $H_y$ field sweep. \textbf{f} The response under external $H_z$ fields.}
    \label{transfercurve_figure}
\end{figure}

To measure an external magnetic field using the SOT effect, the anomalous Hall resistance $R_{xy} = {V_{xy}}/{I_{xx}}$ was determined from the anomalous Hall voltage ($V_{xy}$), which was measured perpendicular to the current flow ($I_{xx}^+$)(see Fig. \ref{cartoon_figure}a) \cite{AnomalousHallEffect}. Since this method is sensitive to the OOP magnetization, the anomalous Hall resistance is proportional to $ m_z $. As mentioned earlier, to eliminate cross-sensitivities, the SOT current direction is reversed, and the subtraction of both resistance values yields the final signal. This approach also has the advantage that any electrical offset present in the measurement is canceled out, resulting in an offset-compensated signal (see Fig. \ref{transfercurve_figure}d,e). This occurs because the offset remains constant for both current directions, leading to a cancellation when subtracting the two cases, which can be seen in Fig.~\ref{transfercurve_figure}a,b. An external $H_x$ field can therefore be determined using Eq. \eqref{analytical_equation_Sx}, represented by the blue data in Fig. \ref{transfercurve_figure}d. By substituting Eq. \eqref{equation_Rxy_prop_mz} into Eq. \eqref{analytical_equation_Sx}, the expression can be rewritten in terms of $R_{xy}$, which is directly measurable.

\begin{equation}
{S}_{x} = \frac{1}{2} \cdot \left( R_{xy} \left( I_{xx}^+ \right) - R_{xy}\left(I_{xx}^-\right) \right)
\label{SOT_signalx_quation}
\end{equation}

Here $R_{xy}\left( I_{xx}^{\pm} \right)$ is the anomalous Hall resistance for positive/negative SOT current direction. Because of symmetry, the same is possible to obtain the signal for external $H_y$ fields, illustrated as the blue curve with squares in Fig. \ref{transfercurve_figure}e.

\begin{equation}
{S}_{y} = \frac{1}{2} \cdot \left( R_{yx} \left( I_{yy}^+ \right) - R_{yx}\left(I_{yy}^-\right) \right)
\label{SOT_signaly_quation}
\end{equation}

In this case the SOT current is applied in y-direction, denoted with $I_{yy}$, and the anomalous Hall voltage measured in x-direction ($V_{yx}$). In the analytical expression \eqref{analytical_equation_Sx}, this case can be evaluated simply by exchanging $x$ with $y$.

For measuring external z-fields, Eq. \eqref{analytical_equation_Sz} can be used, where any cross-sensitivities are canceled out due to the summation of $ m_{z,I+} $ and $ m_{z,I-} $. However, this does not result in offset compensation, which is more challenging, as the signal does not reverse when the direction of the SOT current is changed (see Eq. \eqref{analytical_equation_Sz} and Fig. \ref{transfercurve_figure}c). In other literature like in Ref. \cite{SOTDeviceSensing3DFields} the $H_z$ field was obtained by adding up the AHE resistances for $\pm I_{xx}$ or $\pm I_{yy}$ like in Eq. \eqref{analytical_equation_Sz}, which does not lead to an offset compensation for the measurement of $H_z$ fields. Additionally, in Ref. \cite{SOTLShaped3DSensor}, the measurement of external z-fields lacks offset compensation, and it was noted that the remaining offset was subtracted from the data manually. In contrast, we employ a compensation method similar to the spinning current principle used in commercial Hall sensors \cite{Patent, Munter_spinning_current_tech}

\begin{equation}
{S}_{z} = \frac{1}{2} \cdot \left( \left( \frac{R_{xy} \left( I_{xx}^+ \right) + R_{xy}\left(I_{xx}^-\right)}{2} \right) -   \left( \frac{R_{yx} \left( I_{yy}^+ \right) + R_{yx}\left(I_{yy}^-\right)}{2} \right) \right)
\label{SOT_signalz_quation}
\end{equation}

By summing the anomalous Hall resistances of $I_{xx}^+$ and $I_{xx}^-$ as well as $I_{yy}^+$ and $I_{yy}^-$, instead of subtracting them, eliminates the SOT contributions from the signal. Subsequently, a standard spinning current technique can be applied, by subtracting the two remaining terms.
This allows an offset compensated measurement of external magnetic fields in all 3 directions, which is visible in Fig. \ref{transfercurve_figure}d-f.
The experimental data were acquired using a current density of $j_e = 1.71 \cdot 10^{11} \, \mathrm{A\,m^{-2}}$ ($I_{xx} = 2.7 \, \mathrm{mA}$), under an external magnetic field ranging from $\mu_0 H_{\mathrm{ext}} = -10 \, \mathrm{mT}$...$10 \, \mathrm{mT}$. The reported current density $j_e$ refers specifically to the value within the HM layer, while the value in parentheses indicates the total current applied across the Ta/CoFeB bilayer. Details regarding the calculation of the current density are provided in Sec.~\ref{Determination_of_current_density_in_HM}. The recording time was $t_r = 0.5 \, \mathrm{s}$ for each measurement.

The sensor showed a linear range of  $\pm 0.5 \, \mathrm{mT}$ in the x- and y-direction and $\pm 1 \, \mathrm{mT}$ in the z-direction. The sensitivity or slope ($\kappa_{i}$ in $\mathrm{mA^{-1}}$) can be extracted within the linear range, utilizing the following equations.

\begin{equation}
    \kappa_{i} = \frac{1}{R_s}\frac{d {S}_{i}}{d  H_i }
    \label{sensitivity_formula}
\end{equation}

Here, $ R_s $ represents the anomalous Hall coefficient, which, in our case, is $ R_s = 0.687 \, \Omega $~\cite{AnomalousHallresistivity} (see Sec.~\ref{Determination_of_R_s}). Since the sensitivities were measured in \( \mathrm{VA^{-1}T^{-1}} \), we define 
\( \xi_i = \left( \frac{R_s}{\mu_0} \right) \kappa_i \) to match the commonly used units 
and simplify notation, accepting that both \( \kappa_i \) and \( \xi_i \) represent sensitivity 
in different unit systems. The extracted values of the sensor for the presented sample are \( \xi_{x} = -590 \, \mathrm{VA^{-1}T^{-1}} \) in the x-direction, \( \xi_{y} = -580 \, \mathrm{VA^{-1}T^{-1}} \) in the y-direction, and \( \xi_{z} = 490 \, \mathrm{VA^{-1}T^{-1}} \) in the z-direction. 
These values can be compared to previously reported sensitivities in Ref.~\cite{sabrispaper}.

The offset ($o_i$) performance of the sensor for vanishing field was, $o_x =  36  \, \mathrm{\mu T}$ in the x-direction, $o_y =  50  \, \mathrm{\mu T}$ in the y-direction and $o_z =  37  \, \mathrm{\mu T}$ in the z-direction, without shielding from the Earth's magnetic field.

To validate these results, the experimental data was compared with single-spin simulations, shown as red circles in Fig. \ref{transfercurve_figure}d-f. It can be seen, that the simulation agrees well with the experiment within the linear range. For the cases with external in-plane (IP) fields the simulation deviates slightly for higher external fields. \newline
The used parameters for the simulations were $\gamma = 2.2128 \cdot 10^5 \, \mathrm{m A^{-1}s^{-1}}$, $\alpha = 0.01$, $M_s = \frac{1.2}{\mu_0} \, \mathrm{Am^{-1}}$, $j_e = 1.71 \cdot 10^{11} \, \mathrm{ Am^{-2}}$, $H_k = 3049 \, \mathrm{Am^{-1}}$, $t = 1 \, \mathrm{nm}$, $\eta _{FL}  = 0.0360$ and $\eta _{DL}  = 0.0436$. With these values and Eq. \eqref{pf_factor_Dieter_equation} one gets $p_f = 46898 \, \mathrm{Am^{-1}}$. Unless stated otherwise, these values were used for the simulations.

\section{Extraction of SOT parameters}
\label{label_extraction_of_Sot_parameters}
A key challenge that arose during this process was the lack of knowledge about the SOT parameters ($\eta_{FL}$, $\eta_{DL}$) and the anisotropy field ($H_k$). To determine these values Eqs. \eqref{analytical_equation_Sx} and \eqref{analytical_equation_Sz} can be utilized, which connect these three variables. This approach assumes that the FM layer can be modeled as a single spin system and that the magnetization turns for high enough SOT current with vanishing external field into the negative y-direction (see Appendices~\ref{section_appendix_A} and~\ref{AppendixEquilibriumM}).

From these equations one can derive the following formulas for $\eta_{DL}$ and $\eta_{FL}$

\begin{equation}
   \eta _{FL}  = \frac{{\kappa_{z}  + H_k \kappa_{z}^2 }}{{ \left| p_f \right| \left( {\kappa_{x}^2  + \kappa_{z}^2 } \right)}}
    \label{analytical_solution_etaFL_Dieter_equation}
\end{equation}

\begin{equation}
    \eta _{DL}  = \frac{{ - \kappa_{x}  - H_k \kappa_{x} \kappa_{z} }}{{ \left| p_f \right| \left( {\kappa_{x}^2  + \kappa_{z}^2 } \right)}}
    \label{analytical_solution_etaDL_Dieter_equation}
\end{equation}

which only depend on the slopes $\kappa_{x}$ and $\kappa_{z}$ and the anisotropy field $H_k$. Since the slopes are determined from the experimental data ($\xi_i = \left({R_s}/{\mu_0}\right)\kappa_i$), the only remaining unknown is $H_k$, which can thus be determined.

To calculate the value for the anisotropy field the following scheme was used. Single spin simulations are carried out, with the SOT parameters defined over the analytical equations Eq. \eqref{analytical_solution_etaFL_Dieter_equation} and Eq. \eqref{analytical_solution_etaDL_Dieter_equation}. The absolute difference between the measured and simulated curves within an external field range of $\pm 10 \, \mathrm{mT}$ is minimized by adjusting the $H_k$ value, which also determines the values of $\eta_{FL}$ and $\eta_{DL}$. Because the slope is fixed in the analytical solution, it does not change with a change in $H_k$. The code optimizes the behavior beyond the linear range. This process was performed simultaneously for the data with external $H_x$ and $H_z$ fields, shown in Fig. \ref{transfercurve_figure}d,f. An initial interval was selected for the anisotropy field $ H_k $. The algorithm identified the optimal value with the smallest deviation from the measured signal, by evaluating equidistant points within the initial interval, then creating a smaller interval around this value and repeating the process. This procedure was iterated five times in total. A detailed mathematical formulation of the optimization algorithm can be found in Appendix~\ref{AppendixB}.

Utilizing Eqs. $\eqref{analytical_solution_etaDL_Dieter_equation}$ and $\eqref{analytical_solution_etaFL_Dieter_equation}$, the values for $\eta_{DL}$ and $\eta_{FL}$ were calculated.

\begin{equation}
    \begin{split}
        & \mu_0H_k = 3.83 \, \mathrm{mT} \\
        & \eta _{DL}  = 0.0436\\
        & \eta _{FL}  = 0.0360
    \end{split}
    \label{extracted_values}
\end{equation}

The final outcome of the extraction is illustrated in the simulation results of Fig.\ref{transfercurve_figure} d-f, which show the comparison between simulation and experiment. The strength of the perpendicular magnetic anisotropy essentially results from the balance of the perpendicular interface anisotropy and the IP shape anisotropy.

\section{Magnetization state}
\label{section_ip_magnetization}
\begin{figure}
    \centering
    \includegraphics[width=\linewidth]{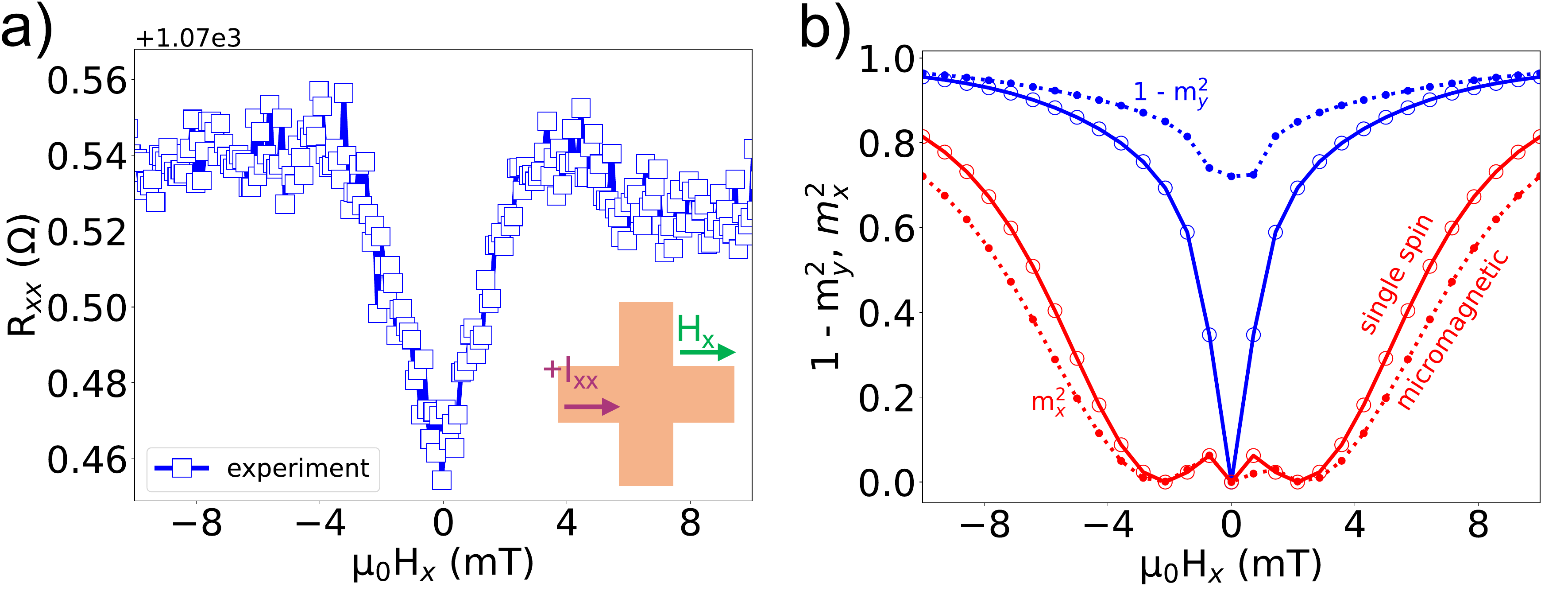}
    \caption{\textbf{a} Measured $R_{xx}$ data over an external $H_x$ field. Visible is the reduced resistance due to the combined AMR and SMR effect around weak external $H_x$ fields. \textbf{b} The result of the single spin simulation (solid lines with circular markers) and the micromagnetic simulation (dotted lines with point markers), of the same situation. Plotted is the AMR effect ($\propto$ $m_x^2$) as the red curve and the SMR effect ($\propto \left(1-m_y^2\right)$) as the blue curve over external $H_x$ field.}
    \label{magnetization_figure}
\end{figure}

In the analytical solution, it was assumed that the magnetization could be approximated as a single spin system. Consequently, analyzing the magnetization state of the FM layer is crucial.
In the experiment, the resistance $R_{xx} = V_{xx}/I_{xx}$ along the current flow was measured (see Fig. \ref{magnetization_figure}). For the measurement a current density of $j_e = 1.84 \cdot 10^{11} \, \mathrm{Am^{-2}}$ ($I_{xx}=2.9 \, \mathrm{mA}$) and an external field between $\pm 10 \, \mathrm{mT}$ was applied in x-direction. The current pulse time was $dt = 0.3 \, \mathrm{s}$ and the recording time was $t_r = 0.25 \, \mathrm{s}$. A clear effect is visible between $\pm 4 \, \mathrm{mT}$ in the measurement, where the $R_{xx}$ values go down from $1070.55 \, \mathrm{\Omega}$ to $1070.46 \, \mathrm{\Omega}$ (see Fig. \ref{magnetization_figure}a). Ref. \cite{Rxx_angle_dependence} reports a change in $R_{xx}$ due to a dependency on the magnetization direction in W/CoFeB/MgO heterostructures, which can be explained mainly by spin Hall magneto resistance (SMR $\propto \left( 1 - m_y^2\right)$). In addition the anisotropic magneto resistance (AMR $\propto m_x^2$) also arises in the system but is reported to be negligible. To examine the magnetization of the FM layer in greater detail, single spin and micromagnetic simulations were performed (see Fig. \ref{magnetization_figure}b). For the micromagnetic simulations, the GPU enhanced finite difference simulation software, magnumnp, was utilized \cite{magnumnp}. The simulation parameters for the micromagnetic simulation were: $A_{ex} = 1.3686 \cdot 10^{-11} \, \mathrm{Jm^{-1}}$, $K_u=5.7479 \cdot 10^5 \, \mathrm{Jm^{-3}}$, $\bm{k} = \left( 0, 0, 1 \right)^\mathrm{T}$ and $D = 0.06 \cdot 10^{-3} \mathrm{Jm^{-2}}$ \cite{DMI_value_paper}. Here, $A_{ex}$ represents the exchange stiffness, $K_u$ the uniaxial anisotropy constant, $\bm{k}$ the anisotropy axis, while $D$ denotes the DMI coefficient. The values for the SOT parameters were taken from the extraction mentioned in \eqref{extracted_values}. The remaining parameters are the same as in the single-spin simulation. The SMR effect is visualized as the blue data, with a solid line and circular markers for the single spin simulation. The dotted lines with point markers denote the micromagnetic simulations. In the simulation, the SMR effect is limited to $\pm 4 \, \mathrm{mT}$ external magnetic field with a sharp peak in case for the single spin simulation. The AMR effect visualized in red has a wider range without a peak for single spin and micromagnetic simulations. The shape and range of the experimental data agrees with the simulated SMR effect and hints to a negligible AMR effect, agreeing with Ref. \cite{Rxx_angle_dependence}. In case for SMR the resistance scales with $1-m_y^2$, leading to lowest resistance values for a maximum $m_y^2$ magnetization and therefore parallel/antiparallel alignment to the spin polarization. The micromagnetic simulation only shows a partial dip in $1 -m_y^2$, not agreeing with the assumption of the analytical solution. Since the correlation between the resistance decrease and the $m_y$ magnetization in the sample is unknown, a definitive conclusion of the magnetization state cannot be drawn.

\begin{figure}[ht]
    \centering
    \includegraphics[width=\linewidth]{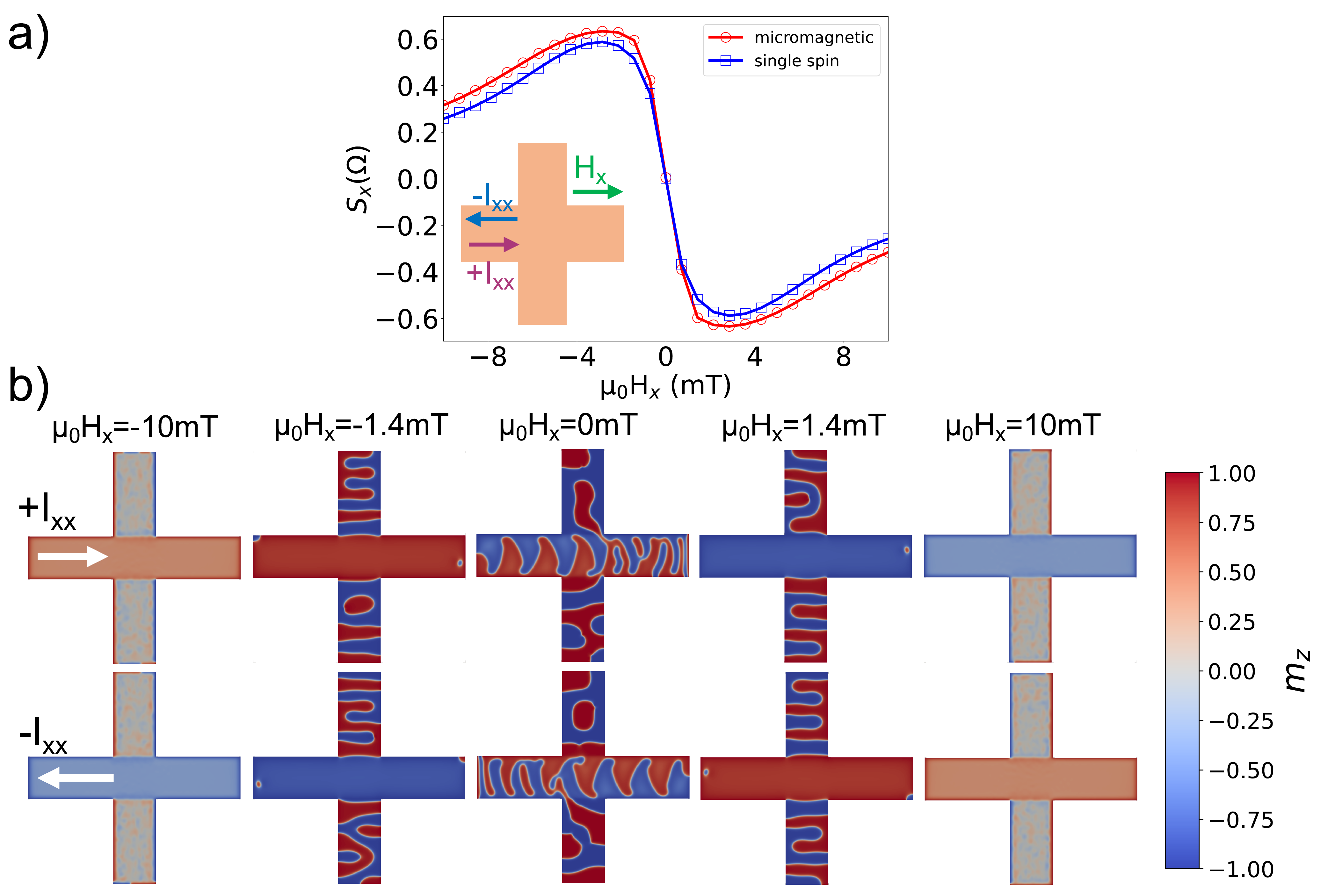}
    \caption{\textbf{a} Comparison of $S_x$ over an external $H_x$ field sweep of the single spin simulation (blue curve) and the micromagnetic simulation (red curve) with an applied SOT current density $j_e= 1.71 \cdot 10^{11} \, \mathrm{Am^{-2}}$. \textbf{b} Snapshots of the micromagnetic simulation over an external $H_x$ field sweep for positive (top row) and negative (bottom row) SOT current. Visualized is the $m_z$ component of the magnetization for the same $H_x$ field sweep. Blue areas denote negative magnetization values and red positive values, which can be seen in the colorbar on the right side.}
    \label{micro_magn_sim_figure}
\end{figure}

To better understand this, we compare the micromagnetic simulation data, performed under an external magnetic field sweep in the x-direction ranging from $\mu_0 H_x = \pm 10 , \mathrm{mT}$, with a single spin simulation. This comparison is illustrated in Fig. \ref{micro_magn_sim_figure}a, where the single spin simulation (blue curve) and the micromagnetic simulation (red curve) are shown. The calculated signal is plotted as a function of the external $H_x$ field. The micromagnetic simulation shows good agreement with the single spin simulation, albeit with a slightly increased amplitude. In Fig. \ref{micro_magn_sim_figure}b snapshots of the $m_z$ magnetization for different $H_x$ field values and different SOT current directions can be seen. Blue areas denote negative magnetization values and red areas positive magnetization values, visualized in the colorbar on the right in Fig. \ref{micro_magn_sim_figure}b. The top row shows snapshots for positive SOT currents, while the bottom row corresponds to negative SOT currents. The cross-sectional views reveal that, at low external fields ($\left| \mu_0 H_x \right| < 1.4 \, \mathrm{mT}$), the magnetization adopts a multidomain OOP state with a limited IP magnetization (also see Fig. \ref{magnetization_figure}b). This behavior deviates from the assumption of a uniform single-spin system and thus does not align with the results of single-spin simulations. The split-up in stripe domains explains the big difference of SMR effect between single spin and micromagnetic simulations (see Fig. \ref{magnetization_figure} b, blue lines).

Nevertheless, in the region $\left| \mu_0 H_x \right| < 1.4 \, \mathrm{mT}$, the signals from both the single-spin and micromagnetic simulations exhibit good agreement. This correspondence arises because the combined effects of SOT and DMI induce domain wall motion that move the domains outwards, effectively resulting in a single-domain state \cite{DMI_SOT_domain_wall_motion}. Only in the limit of vanishing external field do stable, compensating domains, resembling wave-like patterns driven by the combined SOT and DMI effects, persist, nearly canceling each other’s contributions to the signal. Therefore, consistency is observed in this regime as well (see Fig.~\ref{micro_magn_sim_figure}b).

\section{Behavior under bias fields}
\begin{figure}[ht]
    \centering
    \includegraphics[width=\linewidth]{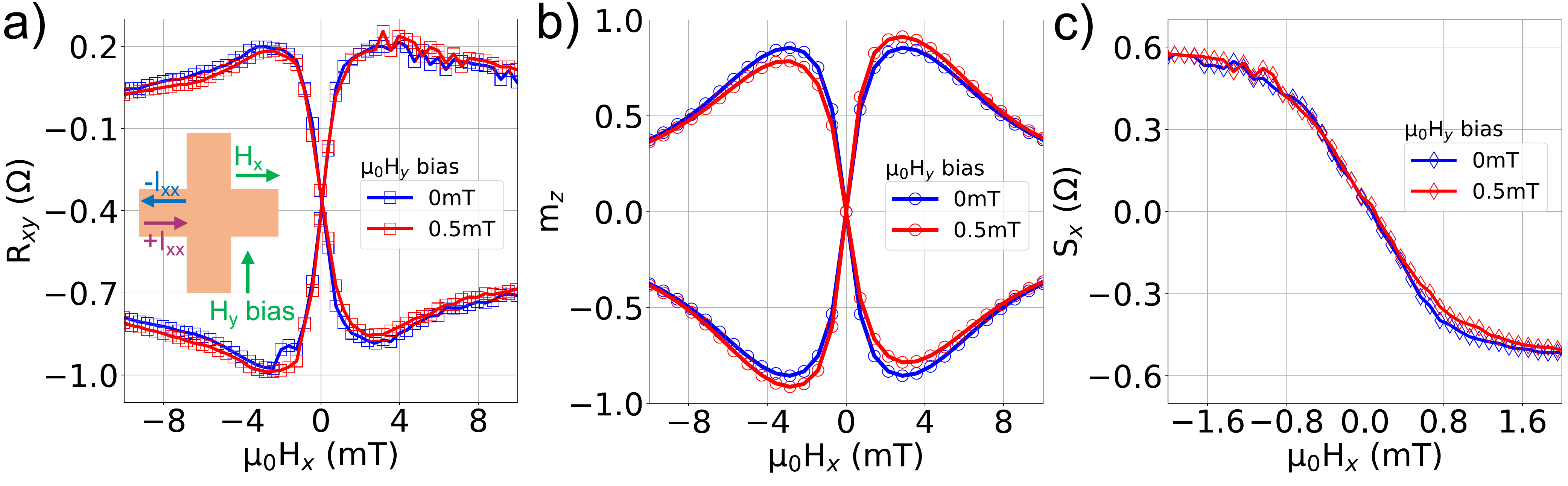}
    \caption{\textbf{a} The measured response curve for an external $H_x$ field sweep, with zero bias field (blue curve with squares) and $\mu_0 H_y = 0.5 \, \mathrm{mT}$ bias field (red curve with squares), for SOT current in $\pm$x-direction. \textbf{b} The simulated data for the same two cases, utilizing a single spin simulation. \textbf{c} The calculated sensor curve for the measured data, using Eq. $\eqref{SOT_signalx_quation}$.}
    \label{bias_field_figure}
\end{figure}

For a 3D magnetic field sensor, the cross sensitivities of all three field components have to be investigated. On the one hand there is the possibility of IP bias fields, while measuring OOP fields. On the other hand there are IP and OOP bias fields, while an IP component (x or y) is measured. Low cross-sensitivity is a prerequisite for the simultaneous measurement of magnetic fields in any arbitrary direction. In Fig. \ref{bias_field_figure}a the case for an IP field (x-field) measurement with an IP bias field in y-direction is shown. Visualized are the measured anomalous Hall resistances $R_{xy}(I_{xx}^+)$ and $R_{xy}(I_{xx}^-)$, for the cases without (blue data) and with bias field ($\mu_0 H_y = 0.5 \, \mathrm{mT}$, red data). A current density $ j_e=\pm 1.71 \cdot 10^{11} \, \mathrm{Am^{-2}}$ ($I_{xx}=\pm2.7 \, \mathrm{mA}$) was applied for a pulse length $dt = 0.6 \, \mathrm{s}$. The recording time was set to $t_r = 0.5 \, \mathrm{s}$ and the external $\mu_0 H_x$ field was swept from $\mu_0 H_{ext}=10 \, \mathrm{mT}$ to $-10 \, \mathrm{mT}$.

The data measured under the influence of a transverse bias field in the y-direction reveal an increased amplitude of the AHE signal for one SOT current polarity and a decreased amplitude for the opposite polarity. This asymmetry arises from the interplay between the field-like torque and the transverse bias field, which together influence the equilibrium configuration of the magnetization. As shown in Eq.~\eqref{analytical_solution_+Ixx_mz_Dieter_equation} in Appendix~\ref{AppendixB}, the OOP magnetization component \( m_z \) in the presence of an x-directed external field is inversely proportional to a denominator that depends on the difference between the field-like torque and the applied bias field \( H_y \). When the bias field and the field-like torque act in the same direction, their combined effect reduces the denominator, enhancing \( m_z \) and thereby increasing the AHE signal. Conversely, if they partially cancel, the denominator becomes larger, reducing the effective canting and leading to a weaker AHE response.

This can be seen in the red curve of Fig. \ref{bias_field_figure}a. For the signal, starting on the left top part, the amplitude of the red curve is slightly lower compared to the blue curve. The opposite happens for the signal with opposite current direction, where the amplitude is increased.

For comparison a single spin simulation of the system was performed, which is visible in Fig. \ref{bias_field_figure}b. The z-magnetization is visualized over an external $H_x$ field sweep, for the same two cases. Without bias field (blue curve with circles) and with $\mu_0 H_y=0.5 \, \mathrm{mT}$ bias field (red curve with circles). Also in the simulation, the amplitude increases for one current direction and decreases for the other direction, like in the experiment.

In Fig. \ref{bias_field_figure}c, the signal (${S}_x$) is plotted as a function of the external $H_x$ field. The cases with no bias field (blue curve with diamonds) and with a bias field of $\mu_0 H_y = 0.5 \, \mathrm{mT}$ (red curve with diamonds) are shown. Both curves exhibit nearly identical sensitivities (slopes), indicating robustness to magnetic fields that are not aligned with the measurement direction. This can be explained by the increasing amplitude for one current direction and the decreasing amplitude for the other current direction in the measured anomalous Hall voltage, which almost perfectly cancels out the effect of external $H_y$ bias fields. The offset changes from $o_x(\mu_0 H_y = 0 \, \mathrm{mT})=50 \, \mathrm{\mu T}$ to $o_x(\mu_0 H_y = 0.5\, \mathrm{mT})=76 \, \mathrm{\mu T}$ in the presence of an external bias field in y-direction. The sensitivity changes from $\xi_{x}(\mu_0 H_y = 0 \, \mathrm{mT}) =583 \, \mathrm{VA^{-1}T^{-1}}$ to $\xi_{x}(\mu_0 H_y = 0.5\, \mathrm{mT}) =557 \, \mathrm{VA^{-1}T^{-1}}$.

\begin{figure}[ht]
    \centering
    \includegraphics[width=\linewidth]{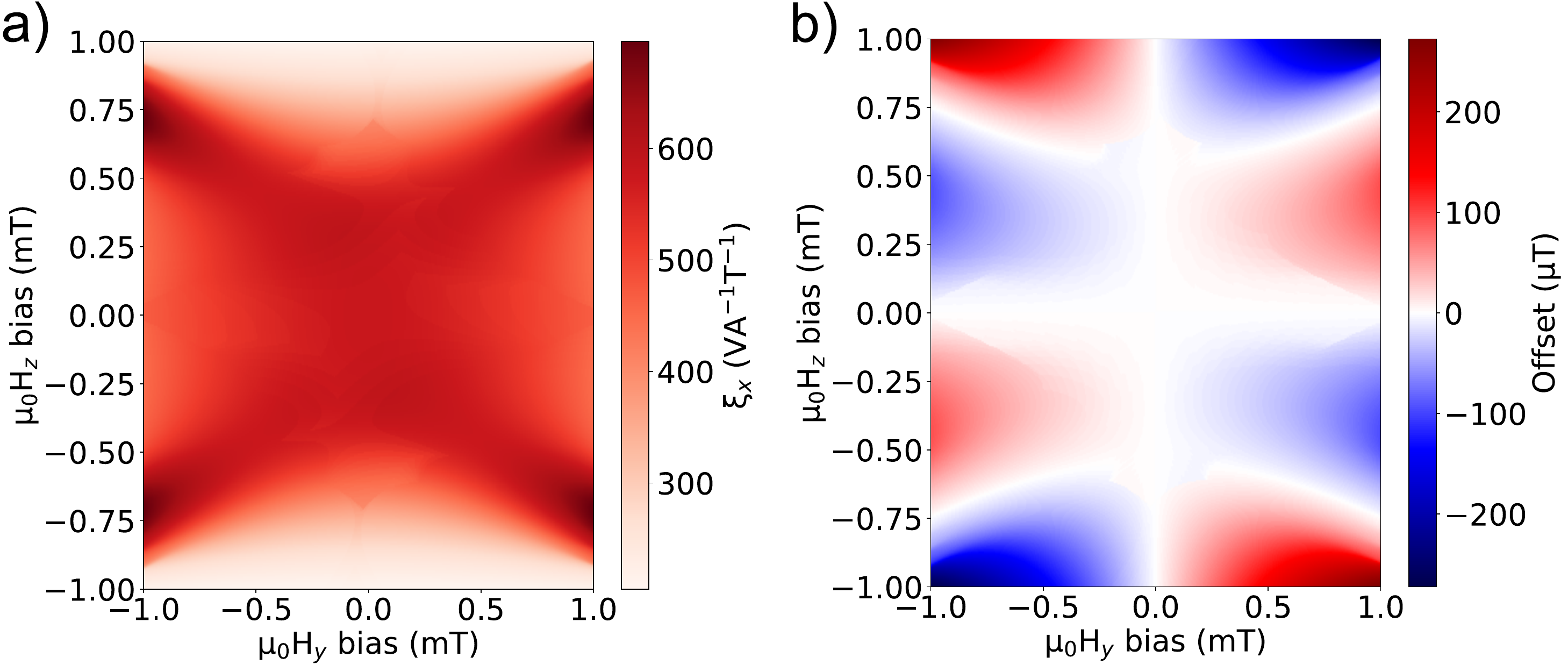}
    \caption{\textbf{a} The simulated sensitivity ($\xi_x$) in $\mathrm{VA^{-1}T^{-1}}$ over external $H_x$ fields for different external bias fields in y- and z-direction utilizing single spin simulations. \textbf{b} The extracted offset values for different external bias fields in y- and z-direction.}
    \label{bias_field_simulation_figure}
\end{figure}

For bias fields in the y- and z-directions, single spin simulations were performed over various amplitudes up to $\pm 1 \, \mathrm{mT}$ per direction, as shown in Fig. \ref{bias_field_simulation_figure}. For every $H_y$ and $H_z$ bias field, a full $H_x$ sweep was performed, like visible in Fig. \ref{bias_field_figure}. This allowed the extraction of the sensitivity and offset data. The resulting sensitivities and offsets naturally depend on the strength of the SOT, determined by the SOT current and associated parameters, and in this case correspond to the experimental conditions shown in Fig. \ref{transfercurve_figure}. In the case where both \( H_y \) and \( H_z \) bias fields were applied simultaneously, only simulations could be conducted, as the magnetic coils were limited to generating a 2D magnetic field. In Fig. \ref{bias_field_simulation_figure}a the data of the sensitivity $\xi_{x} = \left( R_s/\mu_0\right) \kappa_x$ is visualized. The color code defines the amplitude of the data. On the x-axis the $H_y$ field is varied and on the y-axis the $H_z$ field. As visible, a $H_y$ bias field does not change the slope a lot, which agrees with the measurement in Fig. \ref{bias_field_figure}c. For higher $ H_y $ bias fields, the sensitivity decreases from $ \xi_{x} = 577 \, \mathrm{VA^{-1}T^{-1}} $ in the absence of a bias field to $ \xi_{x} = 538 \, \mathrm{VA^{-1}T^{-1}} $ at $ \mu_0 H_y = 0.5 \, \mathrm{mT} $. This reduction occurs because, as the external bias field increases, the SOT effect becomes less significant in comparison. As a result the magnetization of the FM layer will align with the external magnetic field and therefore decrease the SOT effect for both current directions. This will also lead to a decrease of the sensitivity, which can be observed in Fig. \ref{bias_field_simulation_figure}a. This becomes noticeable above the linear range of the sensor, which is around $\pm 0.5 \, \mathrm{mT}$. For the case of external $\mu_0 H_z$ fields the sensitivity drops significantly at bias fields just above $0.5 \, \mathrm{mT}$ from $ \xi_{x} = 577 \, \mathrm{VA^{-1}T^{-1}} $ to $\xi_{x} = 301$ $\mathrm{VA^{-1}T^{-1}}$ for $\mu_0 H_z= 0.8 \mathrm{mT}$.

In Fig. \ref{bias_field_simulation_figure}b the simulated offset values, for different bias fields can be seen. When the bias field is applied in only one direction (either the y- or z-direction), the offset remains zero. However, when the bias field has both y- and z-components, the offset becomes nonzero. For smaller external bias fields up to $ \pm 0.5 \, \mathrm{mT} $ (within the linear range), the offset remains relatively small ($< 30 \, \mathrm{\mu T}$). Beyond this range, the magnetization response to external fields is not linear anymore and therefore can not be utilized for detecting these fields. Despite this the simulation can still be performed successfully, and it appears that the offset begins to decrease again in the region from \(\mu_0 H_y = 0.5\, \mathrm{mT}\) to \(\mu_0 H_z = 0.75\, \mathrm{mT}\), suggesting the onset of some compensation.

\section{Discussion and Conclusion}
A 3D magnetic field sensor with offset compensation in all 3 dimensions has been shown. Utilizing a modulation principle of the magnetization via SOT, an offset performance in the order of tens of $\mu T$ was achieved. For the measurement of the x-component of an external magnetic field with small (within the linear range of the sensor) $H_y$ bias fields, no cross sensitivities were measured. A novel approach has been introduced to also measure the $H_z$ component without offset, employing a technique similar to the conventional spinning current method in Hall effect sensors \cite{Munter_spinning_current_tech}.

The sensitive element is able to detect external magnetic fields in all 3 directions, without major cross sensitivities or changes in the offset, within a certain field range. If the amplitude of the bias field exceeds the linear range of the sensor ($>0.5 \, \mathrm{mT}$), then there is a noticeable change in the sensitivity and therefore in the signal.

The magnetization state of the FM layer could not be fully determined experimentally. Attempts to capture MOKE images of the structure were not successful due to intransparent layers on top of the sensitive structure. Consequently, the analytical extraction of the SOT parameters should be interpreted with caution, as it relies on the assumption of a single spin system. Notably, the signs of the SOT parameters are the same, in contrast to the literature (see page 31 in Ref. \cite{TorqueTerm}), which reports opposite signs. Despite these limitations, the observed behavior was well reproduced by single-spin simulations. This method is expected to perform optimally for an IP FM layer consisting of a single domain and should inspire researchers to extract the SOT parameters ($\eta_{FL}$ and $\eta_{DL}$) from measurements, utilizing this approach.

\section{Methods}
\label{section_methods}
\subsection{\label{Sample_fabrication}Sample fabrication}
The samples examined in this work were produced on the same wafer with the same structure as those described in 
Ref. \cite{sabrispaper}. For the sake of completion, the production method is again mentioned in this work.

The SOT structure was produced on a 8" silicon wafer using a Singulus Rotaris system. Prior to depositing the 3 layer film stack, an aluminum metallization layer was added to the wafer to establish electrical contact with the SOT structure from below. The aluminum layer was connected to the SOT structure through tungsten vias, separated by a $\mathrm{SiO_2}$ insulating layer. A chemical mechanical polishing (CMP) step was performed to smooth the surface before proceeding with the deposition of the SOT film stack. The stack consisted of Ta ($6 \, \mathrm{nm}$)/$\mathrm{Co_{60}Fe_{20}B_{20}}$ ($1 \, \mathrm{nm}$)/MgO ($1.5 \, \mathrm{nm}$)/Ta ($5 \, \mathrm{nm}$) and was deposited using physical vapor deposition (PVD) in a vacuum environment, with a base pressure of less than $5 \cdot 10^{-9} \, \mathrm{torr}$, ensuring no vacuum breaks between layers. Argon served as the sputtering gas for all the layers, with metal layers deposited in DC mode and the MgO layer in RF mode. After deposition, the wafers were annealed in a nitrogen atmosphere for 2 hours at $280 \mathrm{^\circ C}$. 
The Hall bars used in this study were patterned from the films using conventional optical lithography and Ar ion etching. To protect the SOT structure from corrosion, a passivation layer was deposited. Further encapsulation, involving aluminum-filled vias around and metallization layers on top of the SOT structure, enhances heat dissipation. Finally, the pads were released by opening the passivation layer, and standard mechanical dicing techniques were used for separation.

\subsection{\label{Determination_of_R_s}Determination of $R_s$}
To convert resistance values into the normalized magnetization in the  z -direction ($ m_z $), the anomalous Hall resistivity $ \rho_{xy} $ can be used \cite{AnomalousHallresistivity}.

\begin{equation}
    \rho_{xy} = R_0 H_z + R_s M_z
\end{equation}

$R_0 H_z$ represents the ordinary Hall resistivity, with $R_0$ denoting the ordinary Hall coefficient and $R_s M_z$ the anomalous Hall resistivity, with $R_s$ denoting the anomalous Hall coefficient. Because the ordinary Hall effect is almost always negligible compared to the AHE, the measured resistance $R_{xy}$, in the experiment, can be linearly correlated with the magnetization \cite{AnomalousHallresistivity2}:

\begin{equation}
    R_{xy} \approx R_s m_z
    \label{equation_Rxy_prop_mz}
\end{equation}

The AHE coefficient $R_s$ was determined via the AHE resistance $R_{xy}$ for an in z-direction saturated sample.

For the calculation of the AHE coefficient the data of the $H_z$ measurement was used, which can be seen in Fig. \ref{transfercurve_figure}f. The amplitude of ${S}_{z}$, in units of $\Omega$ at highest applied external magnetic field was used (${S}_{z} \left( \mu_0 H_z= \pm 40 \, \mathrm{mT} \right)$). This value was defined equivalent with $m_z=1$ for a current density $j_e = 1.71 \cdot 10^{11} \,\mathrm{Am^{-2}}$ ($I_{xx} = 2.7 \, \mathrm{mA}$), which leads to the AHE coefficient $R_s$.

\begin{equation}
    R_s = 0.5 \cdot \left[ {S}_{z} \left( \mu_0 H_z=40 \, \mathrm{mT} \right) - {S}_{z} \left( \mu_0 H_z=-40 \, \mathrm{mT} \right) \right] = 0.687 \Omega
    \label{AHE_coefficient}
\end{equation}

This allows the conversion from resistance values in units of $\mathrm{\Omega}$ to unit magnetization, which enables the direct comparison between experimental results and simulations, visible in Fig. \ref{transfercurve_figure}d-f.

\subsection{\label{Measurement_equipment}Measurement equipment}
For the measurements the following equipment was used. To apply external magnetic fields, 2D coils from the company Brockhaus Messtechnik GmbH $\&$ Co. KG were used, powered by 2 Kepco power supply units. The power supply and measurement units for the test samples were provided by National Instruments. A PXIe-4139 SMU card was used as the power supply unit, while a PXIe-6366 card with a resolution of \( 2 \, \mathrm{MS/s} \) served as the measurement unit. The entire experimental setup was connected to a computer, which controlled all operations. LabVIEW (National Instruments) was used to measure the samples and send commands to the power supply units. All measurements in this work were conducted using pulsed DC currents. For each value of the external magnetic field, both positive and negative SOT current pulses were applied. After sweeping through both current directions, the external field was incremented, starting and ending with a positive value. During each pulse, a DC current was applied to the test sample, and the measurement was performed. A short rest period followed each pulse. To minimize disturbances from time-varying magnetic fields, the external field was held constant throughout each pulse sequence. 

\subsection{\label{Determination_of_current_density_in_HM}Determination of current density in HM} 
Since only the applied current in the Ta/CoFeB (HM/FM) bilayer is known, the current density within the heavy metal (HM) layer, relevant for the SOT effect, must be calculated. As the individual layer resistivities were not measured, literature values are used for this purpose. For the Ta layer, which is in the $\beta$-phase, Ref.~\cite{beta_ta_resi} reports a resistivity of approximately $\rho_{\mathrm{Ta}} = 200 \, \mathrm{\mu \Omega \cdot cm}$ at a temperature of $T = 300 \, \mathrm{K}$. Considering the sample heating due to high current densities, we assume a slightly reduced resistivity of $\rho_{\mathrm{Ta}} = 190 \, \mathrm{\mu \Omega \cdot cm}$. For the CoFeB layer, Ref.~\cite{beta_ta_resi} also provides a resistivity value of $\rho_{\mathrm{CoFeB}} = 100 \, \mathrm{\mu \Omega \cdot cm}$. Using these two values and a two resistors model, we get the two resistances for the two individual layers with the following equation

\begin{equation}
R = \frac{\rho \cdot L}{w \cdot t}
\label{reistance_from_resistivity}
\end{equation}

Here $\rho$ is the sheet resistivity, $L$ the length of the Hall bar the current is traveling through, $w$ the width of the Hall bar and $t$ the layer thickness. We have the following values: $L = 10 \, \mathrm{\mu m}$, $w = 2 \, \mathrm{\mu m}$, $t_{Ta} = 6 \, \mathrm{nm}$ and $t_{CoFeB} = 1 \, \mathrm{nm}$. The resistivities were already mentioned above. With Eq. \eqref{reistance_from_resistivity} one gets $R_{Ta } = 1583 \, \mathrm{\Omega}$ and $R_{CoFeB} = 5000 \, \mathrm{\Omega}$, which leads to a total resistance of $R_{total} = 1203 \, \mathrm{\Omega}$. This is close to the measured resistance of $R_{xx} = 1070 \, \mathrm{\Omega}$, visible in Fig. \ref{magnetization_figure} a. Assuming ideal parallel conduction with no interfacial resistance between the layers, one gets the individual currents flowing through the layers via the applied total current. In this case we assume a total applied current of $I_{xx} = 2.7 \, \mathrm{mA}$ which was applied in the measurements, visible in Fig. \ref{transfercurve_figure}. This leads to: $I_{Ta} = 2.05 \, \mathrm{mA}$ and $I_{CoFeB} = 0.65 \, \mathrm{mA}$, with the current densities following from these values: $j_{Ta} = 1.71 \cdot 10^{11} \, \mathrm{Am^{-2}}$ and $j_{CoFeB} = 3.25 \cdot 10^{11} \, \mathrm{Am^{-2}}$. This means for the case of an applied current of $I_{xx} = 2.7 \, \mathrm{mA}$ we have a current density of $j_{Ta} = 1.71 \cdot 10^{11} \, \mathrm{Am^{-2}}$ flowing through the HM layer. Because the total resistance does not agree perfectly with the measured resistance the correct values could be a bit off and lead to slightly different SOT parameters (see $\eta_{DL}$ and $\eta_{FL}$ in Eq. \eqref{extracted_values}). However, the exact value does not impact the validity of the results because the extracted SOT parameters, \( \eta_{\mathrm{DL}} \) and \( \eta_{\mathrm{FL}} \), scale inversely with the applied current density ($p_f$ factor in Eqs.\eqref{app_analytical_solution_etaFL_Dieter_equation} and \eqref{app_analytical_solution_etaDL_Dieter_equation} in Appendix~\ref{section_appendix_A}). Consequently, any variation in the simulated current density is exactly offset by a proportional adjustment in the extracted parameters. Put differently, doubling the current density would simply halve the SOT parameters, and vice versa. Therefore, the absolute value of the current density used in the simulation acts merely as a scaling factor and does not affect the fundamental physics or the interpretation of the extracted SOT parameters. It reflects the experimental situation of the data used for the extraction, which can be seen in Fig. \ref{transfercurve_figure}.

\begin{acknowledgments}
This project received funding from Infineon Technologies Austria AG and the FWF project No.~10.55776/I6267. We gratefully acknowledge M.~Kirsch and K.~Pr\"ugl for their valuable assistance with sample fabrication.

Sebastian Zeilinger, F.~Bruckner, S.~Koraltan, J.~M.~Salazar-Mej\'{i}a, and D.~Suess developed and improved the single-spin and micromagnetic codes. Sebastian Zeilinger and D.~Suess derived the analytical solutions. Sebastian Zeilinger, A.~Satz, and J.~G\"uttinger performed measurements at Infineon Technologies. Sebastian Zeilinger, B.~Aichner, P.~Heinrich, and Sophie Zeilinger carried out measurements at the University of Vienna. H.~Br\"uckl conducted measurements at the University for Continuing Education Krems. J.~G\"uttinger, A.~Satz, and D.~Suess supervised the project.

Sebastian Zeilinger prepared the initial manuscript, and all authors contributed to the final version.
\end{acknowledgments}

\begin{appendices}
\section{Analytical solution}
\label{section_appendix_A}
The magnetization measured and simulated in this work can be described with an analytical equation, 
which will be derived in the following part. This approach assumes that the FM layer can be modeled as a single spin system. If the spins are polarized in the y-direction, $\bm{p}$ can be represented in the following way

\begin{align}
\boldsymbol{p} = \begin{pmatrix}
0 \\
1 \\
0
\end{pmatrix}
\end{align}

The total effective field ($\bm{H}^{\mathrm{eff,total}}$), including the DL- and FL-field, can be expressed as follows

\begin{equation}
	\bm{H}^{\mathrm{eff,total}} = \bm{H}^{\mathrm{eff}} - H_{FL}\bm{p}-H_{DL}\bm{m} \times \bm{p}
\end{equation}

With $\bm{H}^{\mathrm{eff}}$ defined as

\begin{equation}
	\bm{H}^{\mathrm{eff}} = \bm{H} + \bm{H}_k
\end{equation}

Here, $\mathbf{H}$ denotes the contribution from external fields, and $\mathbf{H}_k$ is the effective anisotropy field, which in our case always points along the $z$-direction. The anisotropy field is defined as follows

\begin{equation}
	\bm{H}_k = H_k \left( \bm{m} \cdot \bm{k} \right) \bm{k}
\end{equation}

With the anisotropy axis $\bm{k}$ pointing in our case in z-direction \cite{highcurrentSOT}. With the amplitude

\begin{equation}
    H_k = \frac{2 K_1}{\mu_0 M_s}
\end{equation}

Here $M_s$ is the saturation magnetization and $K_1$ the anisotropy constant.
In equilibrium the total effective field has to be parallel to the magnetization (see Eq. \eqref{effective_field_equation}). In other words $\bm{H}^{\mathrm{eff,total}} = g \cdot \bm{m}$, which leads to
\begin{equation}
    \begin{split}
        & \quad H_x + m_z \cdot H_{{DL}} = g \cdot m_x \\
        & \quad H_y - H_{{FL}}  = g \cdot m_y \\
        & \quad H_z - m_x \cdot H_{{DL}} + m_z \cdot H_k = g \cdot m_z
    \end{split}
    \label{equilibirum_H_Dieter_equation}
\end{equation}

Here $m_i$ is the i-th component of the magnetization unit vector and $g$ a proportionality factor. Reformulating this equation to $m_z$ one is left with the following expression

\begin{equation}
	m_z  = \frac{{m_y H_z \left( {H_y  - H_{FL} } \right) - m_y ^2 H_x H_{DL} }}{{\left( {H_y  - H_{FL} } \right)  \left( {H_y  - H_{FL}  - H_k m_y } \right) + m_y ^2 H_{DL} ^2 }}
	\label{mz_analytical_general_case}
\end{equation}

It is possible to replace $H_{DL}$ and $H_{FL}$, utilizing Eq. \eqref{analytical_HDL_Dieter_equation}

\begin{equation}
	m_z  = \frac{{m_y H_z \left( {H_y  - \left| {p_f } \right|\eta _{FL} } \right) - m_y ^2 H_x \left| {p_f } \right|\eta _{DL} }}{{\left( {H_y  - \left| {p_f } \right|\eta _{FL} } \right)  \left( {H_y  - \left| {p_f } \right|\eta _{FL}  - H_k m_y } \right) + m_y ^2 \left| {p_f } \right|^2 \eta _{DL} ^2 }}
	\label{m_z_with_my}
\end{equation}

For small Gilbert damping and strong enough positive current ($I_{xx}^+$), we assume values for $\eta_{DL}$ and $\eta_{FL}$ so that the magnetization almost points completely in the negative y-direction \cite{Dieter_field_free_switching}.
\begin{align}
\boldsymbol{m} = \begin{pmatrix}
m_x \\
-1 \\
m_z
\end{pmatrix}
\end{align}

A more detailed analysis of the equilibrium magnetization similar to Ref.~\cite{Dieter_field_free_switching} can be seen in Appendix~\ref{AppendixEquilibriumM}. This leaves us with the following expression for the z-magnetization for positive SOT current

\begin{equation}
    m_{z,I+} = \frac{- H_x \left| p_f \right| \eta_{DL}  + H_z (\left| p_f \right| \eta_{FL} - H_y)}{{\left( \left| p_f     \right| \eta_{DL} \right)}^2 + (\left| p_f \right| \eta_{FL} - H_y) (\left| p_f \right| \eta_{FL} - H_k - H_y)}
    \label{analytical_solution_+Ixx_mz_Dieter_equation}
\end{equation}

For negative current the magnetization will point in positive y-direction and also the signs of $\left| p_f \right| \eta_{DL/FL}$ change. This leaves us with

\begin{equation}
    m_{z,I-} =  \frac{H_x \left| p_f \right| \eta_{DL}  + H_z (\left| p_f \right| \eta_{FL} + H_y)}{{\left( \left| p_f \right| \eta_{DL} \right)}^2 + (\left| p_f \right| \eta_{FL} + H_y) (\left| p_f \right| \eta_{FL} - H_k + H_y)}
    \label{analytical_solution-Ixx_mz_Dieter_equation}
\end{equation}

In one equation condensed, this looks like

\begin{equation}
    m_{z,I\pm} =  \frac{\mp H_x \left| p_f \right| \eta_{DL}  + H_z (\left| p_f \right| \eta_{FL} \mp H_y)}{{\left( \left| p_f \right| \eta_{DL} \right)}^2 + (\left| p_f \right| \eta_{FL} \mp H_y) (\left| p_f \right| \eta_{FL} - H_k \mp H_y)}
    \label{analytical_solution_pmIxx_mz_Dieter_equation}
\end{equation}

Now assume vanishing $H_y$ fields the equation becomes

\begin{equation}
    m_{z,I\pm} =  \frac{\mp H_x \left| p_f \right| \eta_{DL}  + H_z (\left| p_f \right| \eta_{FL} )}{{\left( \left| p_f \right| \eta_{DL} \right)}^2 + (\left| p_f \right| \eta_{FL} ) (\left| p_f \right| \eta_{FL} - H_k )}
\end{equation}

By subtracting the solution for positive SOT current from that for negative SOT current, one can derive the analytical expression of the signal for external x-fields $S_x$. 

\begin{equation}
   S_{x} \left( H_x \right) = \frac{R_s}{2} \cdot \left( m_{z,I+} - m_{z,I-} \right) = -R_s\frac{\eta_{DL} H_x}{-\eta_{FL} H_k + (\eta_{DL}^2 + \eta_{FL}^2) \left| p_f \right|}
\end{equation}

Similarly, by summing the two cases, the signal $S_z$ for external z-fields can be obtained.

\begin{equation}
    S_{z} \left( H_z \right) = \frac{R_s}{2} \cdot \left(m_{z,I+} + m_{z,I-} \right)= R_s\frac{\eta_{FL} H_z}{-\eta_{FL} H_k + (\eta_{DL}^2 + \eta_{FL}^2) \left| p_f \right|}
\end{equation}

Utilizing the definition of the slopes (Eq. \eqref{sensitivity_formula}) one gets

\begin{equation}
\kappa_{x}  =  - \frac{{\eta _{DL} }}{{ - \eta _{FL} H_k  + \left( {\eta _{DL}^2  + \eta _{FL}^2 } \right)\left| {p_f } \right|}}
\end{equation}

\begin{equation}
\kappa_{z} = \frac{{\eta _{FL} }}{{ - \eta _{FL} H_k  + \left( {\eta _{DL}^2  + \eta _{FL}^2 } \right)\left| {p_f } \right|}}
\end{equation}

Reformulating these equations to the SOT parameters $\eta_{FL}$ and $\eta_{DL}$ one is left with the final equation utilized for the extraction

\begin{equation}
   \eta _{FL}  = \frac{{\kappa_{z}  + H_k \kappa_{z}^2 }}{{ \left| p_f \right| \left( {\kappa_{x}^2  + \kappa_{z}^2 } \right)}}
   \label{app_analytical_solution_etaFL_Dieter_equation}
\end{equation}

\begin{equation}
    \eta _{DL}  = \frac{{ - \kappa_{x}  - H_k \kappa_{x} \kappa_{z} }}{{ \left| p_f \right| \left( {\kappa_{x}^2  + \kappa_{z}^2 } \right)}}
    \label{app_analytical_solution_etaDL_Dieter_equation}
\end{equation}

\section{Equilibrium magnetization}
\label{AppendixEquilibriumM}
Understanding the equilibrium magnetization under varying SOT strength is essential. To this end, we begin with the explicit LLG equation including SOT, as described in Ref.~\cite{extendedLLG}

\begin{align}
\partial_t \bm{m} = & - \frac{\gamma}{1 + \alpha^2} \, \bm{m} \times \left[ \bm{H}_k + \left( \alpha H_{DL} - H_{FL} \right) \bm{p} \right] \nonumber \\
& - \frac{\alpha \gamma}{1 + \alpha^2} \, \bm{m} \times \left( \bm{m} \times \left[ \bm{H}_k - \left( \frac{1}{\alpha} H_{DL} + H_{FL} \right) \bm{p} \right] \right)
\end{align}

To obtain a representation independent of the magnitude of \( H_k \), we introduce relative coordinates \( H_{DL}/H_k \) and \( H_{FL}/H_k \), yielding

\begin{equation}
\begin{split}
    \partial _t \bm{m} = \left( {  \frac{{\gamma H_k }}{{1 + \alpha ^2 }}} \right) \Bigg(- \bm{m} \times \left[ {\hat {\bm{e}}_z m_z + \left( {\alpha \left( {\frac{{H_{DL} }}{{H_k }}} \right) 
    - \left( {\frac{{H_{FL} }}{{H_k }}} \right)} \right)\bm{p}} \right] \\
    - \alpha \bm{m} \times \left( \bm{m} \times \left[ {\hat {\bm{e}}_z m_z - \left( { \frac{1}{\alpha }\left( {\frac{{H_{DL} }}{{H_k }}} \right) 
    + \left( {\frac{{H_{FL} }}{{H_k }}} \right)} \right)\bm{p}} \right] \right) \Bigg)
\end{split}
\end{equation}

Here, \( \hat {\bm{e}}_z = (0, 0, 1) \) denotes the unit vector in the z-direction. We now perform a rescaling of the time $t$ to make both sides of the equation dimensionless. However, caution is required when \( H_k \) changes sign, as this would formally correspond to negative time evolution. To avoid this issue, we separate the sign and magnitude of \( H_k \), treating \( \lvert H_k \rvert \) in the prefactor and keeping the sign explicitly in the dynamics

\begin{equation}
\begin{split}
    \partial _t \bm{m} = \left( {  \frac{{\gamma \left| {H_k } \right| {\mathop{\rm sgn}} \left( {H_k } \right) }}{{1 + \alpha ^2 }}} \right) \Bigg(- \bm{m} \times \left[ {\hat {\bm{e}}_z m_z + \left( {\alpha \left( {\frac{{H_{DL} }}{{H_k }}} \right) 
    - \left( {\frac{{H_{FL} }}{{H_k }}} \right)} \right)\bm{p}} \right] \\
    - \alpha \bm{m} \times \left( \bm{m} \times \left[ {\hat {\bm{e}}_z m_z - \left( {  \frac{1}{\alpha }\left( {\frac{{H_{DL} }}{{H_k }}} \right) 
    + \left( {\frac{{H_{FL} }}{{H_k }}} \right)} \right)\bm{p}} \right] \right) \Bigg)
\end{split}
\end{equation}

We now define the rescaled time \( t' \) via

\begin{equation}
	t' = t \cdot  \left( { \frac{{\gamma \left| {H_k } \right| }}{{1 + \alpha ^2 }}} \right)
\end{equation}

This rescaling allows for a unified treatment of cases where \( H_k \) is positive or negative, while always evolving the system forward in (rescaled) time. The resulting dimensionless equation is

\begin{equation}
\begin{split}
    \partial _{t'} \bm{m} =  -{\mathop{\rm sgn}} \left( {H_k } \right) \cdot  \Bigg( \bm{m} \times \left[ {\hat {\bm{e}}_z m_z + \left( {\alpha \left( {\frac{{H_{DL} }}{{H_k }}} \right) 
    - \left( {\frac{{H_{FL} }}{{H_k }}} \right)} \right)\bm{p}} \right] \\
    + \alpha \bm{m} \times \left( \bm{m} \times \left[ {\hat {\bm{e}}_z m_z - \left( { \frac{1}{\alpha }\left( {\frac{{H_{DL} }}{{H_k }}} \right) 
    + \left( {\frac{{H_{FL} }}{{H_k }}} \right)} \right)\bm{p}} \right] \right) \Bigg)
\end{split}
\label{eq_part_dif_eq_single_spin}
\end{equation}

Since we are only interested in the final equilibrium state and not the full time evolution, the specific value of \( t'_f \) merely determines when the system has relaxed. Equation~\eqref{eq_part_dif_eq_single_spin} is integrated using the vode solver from the SciPy library, configured with the backward differentiation formula (BDF) method, across a range of values for \( H_{DL}/H_k \) and \( H_{FL}/H_k \).

The initial magnetization is set to \( \bm{m}_{\mathrm{init}} = (1, 0, 0) \), which contrasts with Ref.~\cite{Dieter_field_free_switching}, where the initial state pointed along the z -axis. We use a rescaled time step of \( \Delta t' = 0.1 \, \mathrm{s^2 A^2 kg^{-1} m^{-1}} \), and integrate up to \( t'_f = 100 \, \mathrm{s^2 A^2 kg^{-1} m^{-1}} \). The Gilbert damping parameter is fixed at \( \alpha = 0.05 \), and a positive anisotropy field is assumed. The remaining simulation parameters were kept unchanged and are listed at the end of Sec.~\ref{measurement_of_3D_fields}.


\begin{figure}[ht]
    \centering
    \includegraphics[width=\linewidth]{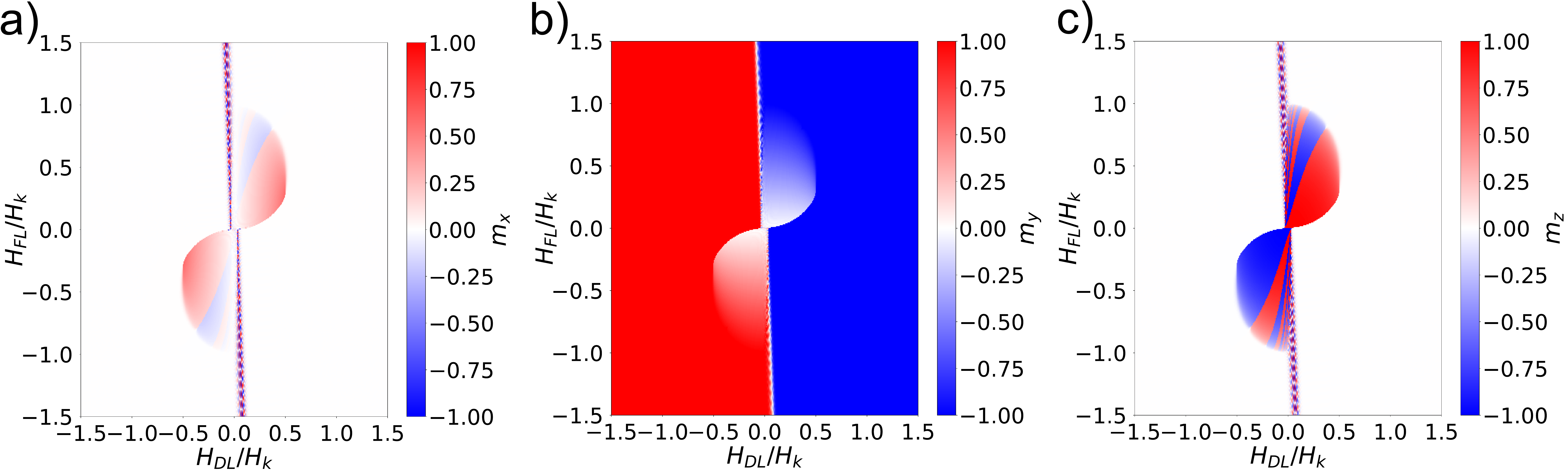}
    \caption{Simulated equilibrium magnetization in the x- (\textbf{a}), y- (\textbf{b}), and z-direction (\textbf{c}) for different SOT strengths, normalized to the anisotropy field. Values range up to \( \pm1.5 \cdot H_{FL/DL}/H_k \).}
    \label{equilibrium_magnetization_direction}
\end{figure}

The results show that, for sufficiently strong SOT $\left(\left|H_{DL/FL}/H_k\right| > 0.5 \right)$, the magnetization predominantly relaxes into the y-direction. For positive values of of the damping-like torque the system relaxes mostly in the negative y-direction agreeing with the assumption of the analytical solution (see Appendix~\ref{section_appendix_A}). The thin, nearly vertical features correspond to persistent magnetization oscillations, indicating non-converging dynamics, consistent with the observations in Ref.~\cite{Dieter_field_free_switching}.


\section{Anisotropy field extraction}
\label{AppendixB}
To determine the optimal anisotropy field \( H_k \), an iterative minimization scheme was employed based on minimizing the absolute error between measured and simulated signals under external fields \( H_x \) and \( H_z \). Since the measured and simulated signals are subtracted from each other, a distinct nomenclature becomes necessary, even though they represent the same quantity.
\subsubsection*{Definitions}

Let:
\begin{itemize}
    \item \( S_x^{\mathrm{meas}}(H_{x,i}) \) and \( S_z^{\mathrm{meas}}(H_{z,i}) \) be the measured signals,
    \item \( S_x^{\mathrm{sim}}(H_{x,i}, H_k) \) and \( S_z^{\mathrm{sim}}(H_{z,i}, H_k) \) be the simulated signals dependent on \( H_k \),
    \item \( \mu_0 H_{x/z,i} \in [-10\,\mathrm{mT}, 10\,\mathrm{mT}] \) be a set of \( N_H \) equidistant external field points.
\end{itemize}

\subsubsection*{Error Function}
The total error for a given $H_k$ is defined as the sum of absolute differences between simulated and measured signals:

\begin{equation}
\begin{split}
\Delta(H_k) = {} & \sum_{i=1}^{N_H} \left( S_x^{\mathrm{meas}}(H_{x,i}) - S_x^{\mathrm{sim}}(H_{x,i}, H_k) \right)^2 \\
& + \sum_{i=1}^{N_H} \left( S_z^{\mathrm{meas}}(H_{z,i}) - S_z^{\mathrm{sim}}(H_{z,i}, H_k) \right)^2
\end{split}
\end{equation}

\subsubsection*{Optimization Procedure}

\begin{enumerate}
    \item Initialize a search interval for \( H_k \): \([H_{k,\mathrm{min}}, H_{k,\mathrm{max}}]\).
    \item Choose the number of discrete evaluation points per iteration, \( N_k \), and total number of refinement steps \( n_{\mathrm{iter}}  \).
    \item For each iteration \( j = 1, \ldots, n_{\mathrm{iter}} \):
    \begin{enumerate}
        \item[3.1] Define \( N_k \) equidistant test points \( H_k^{(1)}, \ldots, H_k^{(N_k)} \) within the current interval.
        \item[3.2] Evaluate \( \Delta(H_k^{(i)}) \) for all \( i = 1, \ldots, N_k \).
        \item[3.4] Identify the minimizing value:
        \[
        H_k^* = \arg\min_{H_k^{(i)}} \Delta(H_k^{(i)})
        \]
        \item[3.4] Define a new narrower interval around \( H_k^* \):
        \[
        H_{k,\mathrm{min}}^{(j+1)} = H_k^* - \delta_j, \quad H_{k,\mathrm{max}}^{(j+1)} = H_k^* + \delta_j
        \]
        with \( \delta_j = \frac{H_{k,\mathrm{max}}^{(j)} - H_{k,\mathrm{min}}^{(j)}}{N_k} \).
    \end{enumerate}
    \item After the final iteration, the optimal anisotropy field is taken as:
    \[
    H_k^{\mathrm{opt}} = H_k^*
    \]
\end{enumerate}

In this work we used $\mu_0H_{k,\mathrm{min}} = 0 \, \mathrm{mT}$, $\mu_0H_{k,\mathrm{max}} = 100\, \mathrm{mT}$, $N_k = 20$ and $n_{iter} = 5$.

\end{appendices}

\input{apssamp.bbl}

\end{document}

%% file: apssamp.bbl
\providecommand{\noopsort}[1]{}\providecommand{\singleletter}[1]{#1}%
%